\documentclass[a4paper,12pt]{article}

\newcommand{\sect}[1]{\setcounter{equation}{0}\section{#1}}

\begin{document}
\topmargin 0pt
\oddsidemargin 0mm

\renewcommand{\thefootnote}{\fnsymbol{footnote}}
\begin{titlepage}
\begin{flushright}
OU-HET 363\\
MCTP-TH-00-17\\
hep-th/0101069
\end{flushright}
\begin{center}
{\Large \bf  OM Theory and V-duality} 
\vspace{2mm} 
\begin{center}
{\large R.-G. Cai\footnote{cai@het.phys.sci.osaka-u.ac.jp},
        J. X. Lu\footnote{jxlu@umich.edu},
N. Ohta\footnote{ohta@phys.sci.osaka-u.ac.jp},
 S. Roy\footnote{roy@tnp.saha.ernet.in}
and Y.-S. Wu\footnote{wu@physics.utah.edu}\\}
\vspace{2mm}

{\em $^\ast$ $^\ddagger$Department of Physics, Osaka University,
 Toyonaka, Osaka 560-0043, Japan\\
$^\dagger$Michigan Center for Theoretical Physics, Randall Physics
 Laboratory\\
 University of Michigan, Ann Arbor MI 48109-1120, USA\\
$^\S$Theory Division, Saha Institute of Nuclear Physics\\
 1/AF Bidhannagar, Calcutta-700 064, India \\
$^\P$ Department of Physics, University of Utah, Salt Lake City, Utah 84112,
   USA}
\end{center}
\end{center}

\vspace{4mm}
\centerline{{\bf{Abstract}}}
\vspace{2mm}
\begin{small}
We show that the (M5, M2, M2$'$, MW) bound state solution of eleven 
dimensional supergravity  recently constructed in
hep-th/0009147 is related to  the (M5, M2) bound state one by a 
finite Lorentz boost along a M5-brane direction perpendicular to the 
M2-brane.  Given the (M5, M2) bound state as a defining system for OM 
theory and the above relation between this system and the 
(M5, M2, M$2'$, MW) bound state, we test the recently proposed V-duality 
conjecture in OM theory. Insisting to have a decoupled OM theory, we
find that the allowed Lorentz boost has to be infinitesimally small, 
therefore resulting in a family of OM theories related by  Galilean boosts. 
We argue that such related OM theories are equivalent to each other.
In other words,  V-duality holds for OM theory as well. Upon compactification
on either an electric or a `magnetic' circle (plus T-dualities as well),
the V-duality for OM theory gives the known one for either 
noncommutative open string theories or noncommutative Yang-Mills
theories. This further implies that V-duality holds in general for 
the little m-theory without gravity.
\end{small}

\end{titlepage}

\newpage
\renewcommand{\thefootnote}{\arabic{footnote}}
\setcounter{footnote}{0}
\setcounter{page}{2}

\sect{Introduction}
Recently there is a surge of interest in studying the effect of a
constant background Neveu-Schwarz (NS)  $B$
field on the decoupled theory of D-branes in superstring theory.
 It has been found that the worldvolume
coordinates of Dp-branes can become noncommutative along the directions
of a nonvanishing  $B$-field.
When the nonzero $B$ field is space-like, one can define a decoupling
limit for a noncommutative Yang-Mills theory (NCYM), i.e., a
noncommutative field theory \cite{SW}. On the other hand, 
if the $B$ field is time-like, a so-called noncommutative open string
(NCOS) theory can be 
achieved \cite{SST,Gop1}.

The counterpart of NS $B$ field in M-theory is a three-form  $C$-field.
Here we have an open M2 brane ending on M5-branes which plays a similar
role as an open string ending on Dp-branes does in superstring 
theory. So a natural question is to ask whether there exists a decoupled
theory in M-theory in which the $C$ field plays 
a similar role as the NS $B$ field does to the decoupled theories of
D-branes in string theory \cite{SW,CHL,Chak,Berg,Kawa,Berm}. 
 Indeed it has been found that a near-critical electric field $C_{012}$
defines a decoupled theory in M-theory, namely OM theory~\cite{Gop2,
Berg2}. The physical picture here is that the electric force due to 
the near-critical electric field $C_{012}$ balances the open membrane tension
such that we end up with a finite proper tension in the decoupling from
the bulk (i.e., sending the eleven-dimensional Planck length to zero).
 Furthermore the compactification of 
OM theory on a magnetic circle yields a (1 + 4)-dimensional
noncommutative Yang-Mills theory (NCYM) with a space-space noncommutativity.
Recall that the (1 + 4)-dimensional NCYM is not renormalizable and 
new degrees of freedom enter when the energy
reaches around $1/g^2_{\rm YM}$. Thus the OM 
theory provides the high energy completion of the NCYM.  On the 
other hand, the compactification of OM theory on an electric circle
gives a (1 + 4)-dimensional NCOS
theory with a space-time noncommutativity.
Thus OM theory provides a description of the NCOS theory when its
coupling is strong.
Therefore the (1 + 4)-dimensional NCYM and NCOS have a unified origin in six 
dimensions: OM theory.  The relation of OM theory to the NCOS is quite 
similar to that between  M theory and IIA superstring theory.

In general, a constant $C_{012}$-field  can be traded to a constant M5
brane worldvolume
3-form field strength\footnote{The non-linear self-duality relation
implies the presence of $H_{345}$ as well. The same should be true for
$C_{345}$.} $H_{012}$. From the M5 brane worldvolume perspective, an M5
brane with a constant $H_{012}$ represents a non-threshold bound state
of this M5 brane with delocalized M2 branes along two of the 5 spatial
directions of the M5 brane. The gravity configuration for this (M5, M2)
bound state was given in \cite{Town}. Recently the gravity configuration
of a more general non-threshold (M5, M2, M$2'$, MW) bound state
 has been constructed in \cite{Berg3} by uplifting 
the known (D4, D2, D$2'$, D0) bound state solution 
of the 10-dimensional type IIA supergravity~\cite{Myer,HO} to 11-dimensions. 
This bound state preserves also 1/2 spacetime supersymmetries just
as the (M5, M2) bound state does. Here MW stands for  M-wave and 
M2 and M2' denote the corresponding delocalized M2 branes, respectively,
along directions orthogonal to each other on the M5 brane worldvolume.

Given that the asymptotic configuration of the (M5, M2, M$2'$, MW) bound
 state is related  to that of the (M5, M2) bound state by a finite
 Lorentz boost and the two bound states preserve the same number of
 spacetime supersymmetry, one must be wondering if the two bound states
are in general related to each other by such a boost. This is not
 obvious at all if one simply examines the corresponding gravity
 configurations of the two bound states.  We will show in the following
section that this is indeed true. Given this result, it is natural for
us to test whether the recently proposed boost-related V-duality, 
first for NCOS in \cite{CW} and then elaborated and extended to more 
general cases in \cite{CLW}, works for OM theory as well. We will 
investigate this in the present paper and the answer is positive.

	This paper is organized as follows: In section 2, we show 
that the (M5, M2, M$2'$, MW) bound state is indeed related to the (M5, M2)
bound state by a finite Lorentz boost along a direction perpendicular 
to the M2 brane. In section 3, we show that insisting a decoupled OM 
theory allows only an infinitesimal Lorentz boost which appears to be 
a Galilean one for the decoupled OM theory. In other words, the allowed
non-trivial spacetime boosts connecting a deformed OM theory to the 
original one are Galilean ones\footnote{There are some subtleties
here when the boost is not perpendicular to the original M2
directions. We will discuss this in \cite{clwone}. The `non-trivial
spacetime boost' here means those which are not in the symmetry group
$SO(1, 2)$ of the original OM theory.}; conjecturally they result in 
physically equivalent theories.  
We will discuss the V-duality for OM theory both from the gravity 
perspective and from the M5 brane worldvolume
perspective.
In section 4, we consider the compactification of the deformed OM
theory on either an `electric' circle or a `magnetic' circle\footnote{
Here the `electric' or `magnetic' circle is defined with respect to the 
original OM theory without the boost. In other words, an `electric'
direction is either the $x^1$ or the $x^2$ direction while a `magnetic'
direction is any of the $x^3, x^4, x^5$ directions. Here the definition
for either `electric' or `magnetic' direction is not perfect since 
we have nonvanishing $C_{034}$. But these are just  convenient names which
will be used later when we consider compactifications of the deformed
OM theory.}. We find that
the V-duality for OM theory gives the corresponding known V-duality for
either NCOS or NCYM discussed in~\cite{CW,CLW}. We conclude this paper in
section 5.

\sect{The (M5, M2, M2$'$, MW) bound state}

In this section we show that the (M5, M2, M2$'$, MW) bound state, recently
constructed in~\cite{Berg3} by uplifting the known (D4, D2, D$2'$, D0)
bound state to 11-dimensions, can also be obtained from the (M5, M2)
bound state by a finite Lorentz boost
along a M5-brane direction perpendicular to the M2-brane.
We start with the supergravity solution of the (M5, M2) 
bound state~\cite{Town}, 
\begin{eqnarray}
\label{eq1}
&& ds^2 = H^{-1/3}h^{-1/3}\left[-dt^2 + (dx^1)^2 + (dx^2)^2 + 
    h\left((dx^3)^2 + (dx^4)^2 + (dx^5)^2\right)\right.\nonumber\\
&&~~~~~~~~~\left. +H(dr^2 +r^2d\Omega^2_4)\right], \nonumber\\
&& l_p^3 C = H^{-1} \sin\alpha dt\wedge dx^1 \wedge dx^2 -H^{-1}h
   \tan\alpha dx^3 \wedge dx^4 \wedge dx^5\nonumber\\
&& F_4 =  3\pi N l_p^3\epsilon_4,
\end{eqnarray}
where $l_p$ is the Planck constant in 11-dimensions, $N$ is the number of
M5-branes in the bound state, $\epsilon_4$ is the volume form of
4-sphere with a unit radius, and the function $h$ and the harmonic
function $H$ are  defined as 
\begin{equation}
\label{eq2}
H=1+\frac{R^3}{\cos\alpha\, r^3}, \ \ \ h^{-1} =H^{-1}\sin^2\alpha
  +\cos^2\alpha,
\end{equation}
with $R^3 = \pi N l_p^3$. 
 
We now Lorentz boost this system along the  $x^5$-direction with 
a boost parameter $\gamma$,
\begin{equation}
\label{eq3}
t \to t \cosh\gamma -x^5 \sinh\gamma, \ \ \  x^5 \to x^5 \cosh\gamma 
-t\sinh\gamma,
\end{equation}
and we end up with
\begin{eqnarray}
\label{eq4}
&& ds^2 = H^{-1/3}h^{-1/3}[-dt^2 +(dx^1)^2 +(dx2)^2 +(dx^5)^2 +h((dx^3)^2 
   +(dx^4)^2)
    \nonumber \\
&& ~~~~~~~ +(h-1)(\cosh\gamma dx^5 -\sinh\gamma dt)^2 +
     H(dr^2 +r^2 d\Omega_4^2)],
    \nonumber \\ 
&& l_p^3 C= H^{-1}(\sin\alpha \cosh\gamma dt\wedge dx^1 \wedge dx^2
    -\sin\alpha \sinh\gamma dx^1 \wedge dx^2 \wedge dx^5 \nonumber \\
&& ~~~~~~~~ +h\tan\alpha \sinh\gamma dt \wedge dx^3 \wedge dx^4 -h\tan\alpha
    \cosh\gamma dx^3 \wedge dx^4 \wedge dx^5),
\end{eqnarray}
where the 4-form field strength $F_4$ as given in (\ref{eq1}) remains
the same.
Defining
\begin{eqnarray}
\label{eq5}
&& \frac{\sin\theta_2}{\cos\theta_1}= \sin\alpha \cosh\gamma, \ \ \ 
   \tan\theta_1 = \sin\alpha \sinh\gamma, \nonumber \\
&& \frac{\sin\theta_1}{\cos\theta_2}=\tan\alpha \sinh\gamma, \ \ \
   \tan\theta_2 = \tan\alpha \cosh\gamma,
\end{eqnarray}
we then have\footnote{Here we obtain the boosted configuration only for 
$\cos\theta_2 \le \cos\theta_1$. The case with $\cos\theta_2 >
\cos\theta_1$ can be obtained by the same boost but now on the (M5, M2)
bound state with M2 branes along $x^3 x^4$ directions.} 
\begin{eqnarray}
\label{eq6}
\cos\alpha =\frac{\cos\theta_2}{\cos\theta_1}, \ \ \
\cosh\gamma = \frac{\sin\theta_2}{\cos\theta_1}\frac{1}{\sin\alpha}, \ \ \
\sinh\gamma =\frac{\sin\theta_1}{\cos\theta_2}\frac{1}{\tan\alpha}.
\end{eqnarray}
Further if we set
\begin{equation}
\label{eq7}
H'=1+ \frac{R^3}{\cos\theta_1\cos\theta_2\, r^3}, \ \ \
h_i^{-1}= H'^{-1}\sin^2\theta_i +\cos^2\theta_i, \ \ i=1,2,
\end{equation}
we can derive 
\begin{equation}
\label{eq8}
H= H'h_1^{-1}, \ \ \  h^{-1}=h_1 h_2^{-1}.
\end{equation} 
Using the above relations, the solution (\ref{eq4}) can be expressed as
\begin{eqnarray}
\label{eq9}
&& ds^2 = H'^{-1/3}h_1^{-1/3}h_2^{-1/3}[-h_1 dt^2 +h_1((dx^1)^2 +(dx^2)^2)
   +h_2((dx^3)^2 +(dx^4)^2) \nonumber \\
&&~~~~~~~~ +(h_2-h_1)(\cosh\gamma dx^5 -\sinh\gamma dt)^2 +H'(dr^2 
   +r^2 d\Omega_4^2)].
\end{eqnarray}
Using the following three identities,
\begin{eqnarray}
\label{eq10}
&& -h_1 + (h_2-h_1) \sinh^2\gamma = -1 +h_1h_2 \sin^2\theta_1\sin^2\theta_2
   (H'^{-1}-1)^2,\nonumber\\
&&h_1 +(h_2-h_1)\cosh^2\gamma = h_1h_2,\nonumber\\
&&-2(h_2-h_1)\cosh\gamma\sinh\gamma = 2h_2h_1(H'^{-1}-1)\sin\theta_1
  \sin\theta_2,
\end{eqnarray}
 we can re-express the boosted 
solution (\ref{eq4}) as
\begin{eqnarray}
\label{eq13}
&& ds^2 = H'^{-1/3}h_1^{-1/3}h_2^{-1/3}[-dt^2 +h_1((dx^1)^2 +(dx^2)^2)
   +h_2((dx^3)^2 +(dx^4)^2) \nonumber \\
&& ~~~~~+ h_1h_2(dx^5 +\sin\theta_1\sin\theta_2(H'^{-1}-1)dt)^2
    +H'(dr^2 +r^2d\Omega_4^2)],
\end{eqnarray}
and
\begin{eqnarray}
\label{eq14}
&& l_p^3 C = H'^{-1}(\frac{\sin\theta_2}{\cos\theta_1}h_1 dt\wedge dx^1
  \wedge dx^2+\frac{\sin\theta_1}{\cos\theta_2}h_2 dt\wedge dx^3 \wedge dx^4
    \nonumber \\
&&~~~~~~~ -h_1\tan\theta_1dx^1\wedge dx^2 \wedge dx^5 
  -h_2\tan\theta_2dx^3 \wedge dx^4
   \wedge dx^5),\nonumber\\
&& F_4 =  3\pi N l_p^3\epsilon_4.
\end{eqnarray}
Eqs.(\ref{eq13}) and (\ref{eq14}) together are nothing but the gravity
 configuration of the (M5, M2, M$2'$, MW) bound state that was recently
 constructed in \cite{Berg3} by uplifting
the (D4, D2, D$2'$, D0) bound state solution  \cite{Myer,HO} to
 11-dimensions.

\sect{The Galilean nature of OM theory: V-duality}

In this section we intend to study a possible OM theory decoupling limit
 for the boosted (M5,M2) system  as given in (\ref{eq4}). 
The usual OM theory decoupling limit as given in \cite{Gop2,Berg2} 
is actually with respect to a static (M5, M2) system with non-vanishing
asymptotic $C_{012}$ and $C_{345}$. The presence of $C_{345}$ originates
 from a non-linear self-duality relation on the M5 brane worldvolume 
3-form field strength \cite{Perry,Howe}. This decoupling limit requires 
a near-critical
 electric field $C_{012}$ whose force almost balances the open membrane
 tension in a limit in which the bulk gravity decouples. As a result,
 the open membrane is confined within the M5 brane worldvolume with a
 finite tension which defines the scale for the OM theory.

 For the boosted (M5,M2) system given
in (\ref{eq4}), one cannot blindly use the boosted electric component 
$C_{012}$ to define the near-critical field limit, since in this case the
net force acting on the open membrane is not merely due to this $C_{012}$
and the other non-vanishing components of the background $C$-field 
such as $C_{034}$ contribute to the net force as well. What we should
account for is the total net force acting on the open membrane. Given
the fact that we are considering a boosted system from the original
static one, we must conclude that the near-critical field limit obtained
for the static case remains the same even for the present boosted
system.
 
	The above indicates that a near-critical field limit is 
independent of a Lorentz boost. This in turn seemingly implies that a
Lorentz boost is allowed for a decoupled OM theory. Our investigation
in \cite{clwone} tells that except for a static configuration, a
near-critical field limit itself is in general not enough to define a 
decoupled theory. In addition, either a well-defined decoupled gravity 
description or a proper open brane description (when available) is needed.
In the following, we first discuss an OM theory decoupling limit for the
boosted system in a gravity setup. We find that insisting a well-defined
OM theory allows only infinitesimal Lorentz boost. This boost appears as 
a Galilean one to the decoupled theory. This indicates that the 
V-duality discussed recently in \cite{CW,CLW} holds for OM theory as
well. We will check this using the proposed open membrane metric 
in~\cite{Berg,Berg2} as well. The conclusion remains the same. 
	
	As mentioned above, the near-critical field limit for the
boosted system remains the same as that for (M5, M2). In other words,
we have now ($\epsilon \to 0$)
\begin{equation}
l_p = \epsilon l_{\rm eff}, \ \ \cos\alpha = \epsilon^{3/2}.
\label{eq:ncl}
\end{equation}
Note that the scalings for the asymptotic transverse coordinates should
remain the same as for the case without boost. This gives $r =
\epsilon^{3/2} u$ with fixed $u$. With these, insisting the decoupling of a
flat bulk region from that of the M5 branes requires further the following 
\begin{equation}
x^{0, 1, 2} = \tilde x^{0, 1, 2}, \ \ x^{3, 4, 5} = \epsilon^{3/2} 
\tilde x^{3, 4, 5}, \  \ \gamma = \tilde v \epsilon^{3/2},
\label{eq:omsl}
\end{equation}
where $\tilde x^\mu$ with $\mu = 0, 1, \cdots 5$ (here $x^0 = t$)
and $\tilde v$ remain fixed. With the above, the gravity description of
OM theory is now  
\begin{eqnarray}
\label{3eq15}
&& ds^2 = \epsilon^2 \left (\frac{u^3}{\pi N l^3_{\rm eff}}\right)^{1/3}
   \tilde{h}^{-1/3} \left [ -d\tilde{t}^2 +(d\tilde{x}^1)^2
   +(d\tilde{x}^2)^2
   + \tilde{h} ((d\tilde{x}^3)^2 + (d\tilde{x}^4)^2)
   \right. \nonumber \\
&& ~~~~~~~~ \left.+ \tilde{h} (d\tilde{x}^5 -\tilde{v} d\tilde{t})^2
   +\frac{\pi N l^3_{\rm eff}}{u^3}\left ( du^2 +u^2 d\Omega_4^2\right)
   \right ], \nonumber \\
&&  C = \frac{ u^3}{\pi N l^6_{\rm eff}}d\tilde{t} \wedge d
    \tilde{x}^1 \wedge d\tilde{x}^2
  -\frac{u^3}{\pi N l^6_{\rm eff}}\tilde{h}d\tilde{x}^3 \wedge
    d\tilde{x}^4 \wedge (d\tilde{x}^5 -\tilde{v} d\tilde{t}),
\end{eqnarray}
where $\tilde{h}^{-1} = 1 +u^3/\pi N l^3_{\rm eff}$.
When $\tilde{v}=0$, the gravity dual reduces to one for the usual
OM theory \cite{Harm}.

	The above results confirm our expectation. A well-defined
decoupled OM theory requires the boost $\gamma = \tilde v \epsilon^{3/2}$
to be infinitesimally small. This appears to be a Galilean boost which 
relates the usual OM theory to the present one through
\begin{equation}
\tilde t \to \tilde t, \ \ \tilde x^5 \to \tilde x^5 - \tilde v 
\tilde t.
\label{eq:gb}
\end{equation}

The background $C$-field (or the M5 brane worldvolume 3-form field
strength $H$) is needed to define the OM theory. The presence of this
field breaks the M5 brane worldvolume Lorentz symmetry. But this
breaking is spontaneous. We therefore expect that the OM theories resulting 
from the respective decouplings in such related backgrounds 
are physically equivalent. Along the same line as discussed in
\cite{CLW}, we have therefore V-duality holding true for OM theory.

For the rest of this section, we make an independent check of what has
been achieved above using the effective open membrane
metric\footnote{This metric was proposed to play a similar role as the
effective open string metric given by Seiberg and
Witten~\cite{SW}. However, unlike the open string metric,
this metric cannot be derived directly due to our inability at present
to quantize the membrane.}(the one seen by the open membrane) 
recently proposed in~\cite{Berg,Berg2}. The nonlinear 
self-duality condition for the M5 brane worldvolume 3-form field
strength $H$  reads \cite{Howe}
\begin{equation}
\label{3eq1}
\frac{\sqrt{-g}}{6}\epsilon_{\mu\nu\rho \sigma \lambda \tau}H^{\sigma
  \lambda \tau}=\frac{1+K}{2}(G^{-1})_{\mu}^{\, \lambda}H_{\nu\rho\lambda},
\end{equation}
where $g_{\mu\nu}$ is the induced metric on the M5 brane,
 $\epsilon^{012345}=1$ and the scalar $K$ and the tensor $G$ are given 
by 
\begin{eqnarray}
\label{3eq2}
&& K=\sqrt{1+\frac{l_p^6}{24}H^2}, \nonumber \\
&& G_{\mu\nu}=\frac{1+K}{2K}\left (g_{\mu\nu} +\frac{l_p^6}{4}H^2_{\mu\nu}
  \right ).
\end{eqnarray}
It is understood that the indices in the above equations are raised or
lowered using the induced metric. It was argued in~\cite{Berg,Berg2}
that the symmetric tensor $G_{\mu\nu}$ is related to  the effective
open membrane metric up to a conformal factor.
With an appropriate choice of the conformal factor, this 
proposed metric for OM theory was shown in \cite{Berg2} to reduce to 
the respective effective open string 
metric upon dimensional reductions.

For the boosted (M5,M2) given in (\ref{eq4}), the asymptotic bulk metric
and the $C$ field along the M5 brane directions are
\begin{eqnarray}
\label{3eq3}
&& ds^2 =-dt^2 + g^2_1 \left((d\tilde x^1)^2 +(d\tilde x^2)^2\right)
 + g^2_2 \left((d\tilde x^3)^2 + (d\tilde x^4)^2\right) + g^2_3 (d\tilde
 x^5)^2, \nonumber \\
&& l_p^3 C = g^2_1 \sin\alpha \cosh\gamma dt\wedge d\tilde x^1 \wedge
 d\tilde x^2 - g^2_1 g_3 \sin\alpha
 \sinh\gamma d\tilde x^1\wedge d\tilde x^2 \wedge d\tilde x^5 \nonumber \\
&& ~~~~~ - g^2_2 g_3 \tan\alpha \cosh\gamma d\tilde x^3 \wedge d\tilde
 x^4 \wedge d\tilde x^5 +
    g^2_2 \tan\alpha \sinh\gamma dt \wedge d\tilde x^3 \wedge d\tilde x^4,
\end{eqnarray}
where we have scaled the coordinates according to the symmetry of system
under consideration.

Recall that the C-field can be traded for the H-field. With this in mind and
from (\ref{3eq3}), we have 
\begin{equation}
\label{3eq5}
G_{\mu\nu}' \equiv  g_{\mu\nu} +\frac{l_p^6}{4}H^2_{\mu\nu}= 
 \left (
\begin{array}{cccccc}
{\cal A} 
 &0&0&0&0& 
{\cal B} \\
0& {\scriptstyle \frac{g_1^2(3+\cos2\alpha)}{4}} &0&0&0&0 \\
0&0& {\scriptstyle \frac{g_1^2(3+\cos2\alpha)}{4}} &0&0&0 \\
0&0&0& {\scriptstyle \frac{g_2^2(2 +\tan^2\alpha)}{2}}&0&0 \\
0&0&0&0& {\scriptstyle \frac{g_2^2(2+\tan^2\alpha)}{2}}&0 \\
{\cal B} &0&0&0&0& {\cal C}
\end{array} \right ) 
\end{equation}
and 
\begin{equation}
\label{3eq6}
l_p^6 H^2 = 6 \sin^2\alpha \tan^2\alpha.
\end{equation}
where 
\begin{eqnarray}
\label{3eq7}
{\cal A} &=& \frac{-2 +\sin^2\alpha \cosh^2\gamma
      +\tan^2\alpha \sinh^2\gamma} {2}, \ \
{\cal B} =  -\frac{g_3(3+\cos2\alpha) \sinh 2\gamma 
 \tan^2\alpha}{8}, \nonumber \\ 
{\cal C} &=& \frac{g_3^2(2+\sin^2\alpha \sinh^2\gamma 
   +\cosh^2\gamma \tan^2\alpha)}{2}.
\end{eqnarray}
Following \cite{Berg2}, we propose the tensor $G'_{\mu\nu}$ to 
be conformally related to the open
membrane metric $\tilde{G}_{\mu\nu}$ as 
\begin{equation}
\label{3eq8}
\tilde{G}_{\mu\nu} =f(l_p^6 H^2) G'_{\mu\nu},
\end{equation}
where $f$ is a function of $l_p^6H^2$ and is introduced such that the open
membrane metric is finite in units of $l^2_p$. However, the precise form
for this function $f$ has not been determined. Fortunately, this does
not prevent us from making the present discussion. 

Since the tensor $G'_{\mu\nu}$ differs from the true open membrane
metric only by a conformal factor, we expect that at least its diagonal 
elements should scale the same way since they survive when the
boost is set to zero. Further inspecting the tensor $G'_{\mu\nu}$, we
find that $G'_{11}, G'_{22}, G'_{33}, G'_{44}$ are independent of the
boost as they should be. With the near-critical field limit
(\ref{eq:ncl}), we have now $g_1 \sim 1, g_2 \sim \epsilon^{3/2}$. These
further imply that the ${\cal A}$ and ${\cal C}$ should be fixed, too.
The former determines that the boost must be $\gamma = \tilde v
\epsilon^{3/2}$ with fixed $\tilde v$, i.e., infinitesimally small. The
latter determines $g_3 \sim \epsilon^{3/2}$. These in turn imply that 
${\cal B}$ is also fixed. If we set from the above the following, 
\begin{equation}
g_1 = 1, \ \ g_2 = g_3 = \epsilon^{3/2}, \ \ \gamma = \tilde v
\epsilon^{3/2},
\end{equation}
we have
\begin{equation}
\label{3eq11}
G'_{\mu\nu}=\frac{1}{2} \left (
\begin{array}{cccccc}
-1 + \tilde{v}^2 &0&0&0&0& -\tilde{v} \\
0&1&0&0&0&0 \\
0&0&1&0&0&0 \\
0&0&0&1&0&0 \\
0&0&0&0&1&0 \\
-\tilde{v} &0&0&0&0& 1
\end{array} \right).
\end{equation}
The above $G'_{\mu\nu}$ says that we can set $f = 2 \epsilon^2$ such
that
\begin{equation}
\label{3eq12}
l_p^{-2} \tilde{G}_{\mu\nu}= 2 l^{-2}_{\rm eff} G'_{\mu\nu}.
\end{equation}
Once again the constant $l_{\rm eff}$ sets the effective scale for 
 the decoupled
OM theory. The above open membrane metric $\tilde G_{\mu\nu}$ is
 related to the one without boost precisely by the Galilean boost
(\ref{eq:gb}) as expected.

\sect{Compactifications of the deformed OM theory}

	In this section, we will show that the V-duality for NCOS and
NCYM discussed in \cite{CW,CLW} can be obtained from the V-duality for OM
theory discussed in the previous section through compactifications and 
T-dualities. As mentioned in \cite{CLW}, the V-duality survives
S-duality there. Since U-dualities for the big M-theory are believed to 
be inherited to the little m-theory without gravity, our study so far
indicates that the U-duality related decoupled theory has V-duality if
the original decoupled theory does. This further enforces our conclusion
made in \cite{CLW} and to be reported in a forthcoming paper
\cite{clwone} that V-duality holds in general in the little
m-theory.

\subsection{Compactification on an electric circle}

In this subsection, we consider first the compactification of the
deformed OM theory on an electric circle (on the direction $\tilde x^2$
for definiteness) following~\cite{Gop2}. This will
give a family of (1 + 4)-dimensional NCOS theories related by V-duality.
Further applying T-duality on, for definiteness, $\tilde x^3$, we end up
with a family of (1 + 3)-dimensional NCOS related by V-duality which was
discussed specifically in~\cite{CW,CLW}. In other words, the V-duality for
OM theory is inherited to its descendant theories as well.

	Consider M-theory on $R^{10} \times S^1$(the circle is in the 
$x^2$ direction) with M5-branes wrapping the circle. Scale all bulk
moduli as in the previous section for OM theory\footnote{For
convenience of the present discussion, we have replaced the scaling
parameter $\epsilon$ used in the previous section by $\epsilon^{1/3}$.}
, i.e.,
\begin{eqnarray}
&&l_p = \epsilon^{1/3} l_{\rm eff}, \ \  \cos\alpha = \epsilon^{1/2}, \
\ \gamma = \tilde v \epsilon^{1/2},\nonumber\\
&&g_{\mu\nu} = \eta_{\mu\nu}~ (\mu,
\nu = 0, 1, 2), \ \ g_{ij} = \epsilon \delta_{ij}~ (i, j = 3, 4, 5),\ \
g_{m n}  = \epsilon \delta_{mn}~ (m, n = {\rm transverse}),\nonumber\\
&& l_p^3C =  \sin\alpha \cosh\gamma d\tilde t\wedge d \tilde x^1 \wedge d\tilde
x^2 - \epsilon^{1/2} \sin\alpha \sinh\gamma d\tilde x^1\wedge d\tilde x^2 
\wedge d\tilde x^5 \nonumber \\
&& ~~~~~ - \epsilon^{3/2} \tan\alpha \cosh\gamma d\tilde x^3 \wedge
d\tilde x^4 \wedge d\tilde x^5 +
    \epsilon \tan\alpha \sinh\gamma d\tilde t \wedge d\tilde x^3 \wedge 
d\tilde x^4, 
\label{eq:comsl}
\end{eqnarray}
where we have kept the cosines and sines as well as $\cosh$'s and
$\sinh$'s in $C$ for later convenience.

	The relation between M-theory and IIA implies that the above
decoupled system is equivalent to a decoupled one of D4 branes
in IIA theory with
\begin{eqnarray}
&&\alpha' = \epsilon \alpha'_{\rm eff}, \ \ g_s = \frac{G^2_o}{\sqrt
\epsilon},\ \ \cos\alpha = \epsilon^{1/2}, \ \ \gamma = \tilde v   
\epsilon^{1/2},\nonumber\\
&&g_{\mu\nu} = \eta_{\mu\nu}~
(\mu, \nu = 0, 1), \ \ g_{ij} = \epsilon \delta_{ij}~ (i, j = 3, 4, 5),\ \
g_{m n}  = \epsilon \delta_{mn}~ (m, n = {\rm transverse}),\nonumber\\
&& 2 \pi \alpha' B =  \sin\alpha \cosh\gamma d\tilde t\wedge d \tilde x^1 
   + \epsilon^{1/2} \sin\alpha \sinh\gamma d\tilde x^1\wedge d\tilde x^5,
\end{eqnarray}
where 
\begin{equation}
\alpha'_{\rm eff} = \frac{l^3_{\rm eff}}{R_2}, \ \ G^2_o =
\left(\frac{R_2}{l_{\rm eff}}\right)^{3/2},
\label{eq:pr}
\end{equation}
with $R_2$ the proper (also the coordinate) radius of the circle.

	The above is nothing but the scaling limit for (1 +
4)-dimensional NCOS. To see the V-duality, let us calculate the open
string moduli using Seiberg-Witten relation~\cite{SW} and we have for
the open string metric
\begin{equation}
G_{\alpha\beta} = \epsilon \left(\begin{array}{ccccc}
				- (1 - \tilde v^2)&0&0&0&- \tilde v\\
                                  0&1&0&0&0\\
                                  0&0&1&0&0\\
                                  0&0&0&1&0\\
                                 -\tilde v&0&0&0&1\end{array}\right),
\label{eq:ncosm}
\end{equation}
and for the nonvanishing noncommutative parameters
\begin{equation}
\Theta^{01} = 2\pi \alpha'_{\rm eff},\ \ \Theta^{15} = - 2\pi
\alpha'_{\rm eff} \tilde v.
\label{eq:ncp}
\end{equation}
In the above, $\alpha, \beta = 0, 1, 3, 4, 5$.

	The above describes a family of (1 + 4)-dimensional NCOS
theories which are related to the one with $\tilde v = 0$ by the
following Galilean boost
\begin{equation}
\tilde t \to \tilde t, \ \ \tilde x^5 \to \tilde x^5 - \tilde v \tilde
t.
\label{eq:ngb}
\end{equation}
In other words, we have the V-duality for (1 + 4)-dimensional NCOS as
discussed in~\cite{CW,CLW} which is now obtained from that for OM theory.

	T-duality of the above along, say, the $\tilde x^3$ direction with the 
coordinate radius $R_3$, will give a family of (1 + 3)-dimensional NCOS 
theories which are related to the one with $\tilde v = 0$ again by
V-duality. The V-duality for this  case was particularly studied
in~\cite{CW,CLW}.
Now we see that this V-duality can also be derived from that for OM
theory. This T-duality leaves the noncommutative parameters
intact and reduces trivially the open string metric in (\ref{eq:ncosm})
to rank 4 through dropping $G_{3\beta}$ and $G_{\alpha 3}$. The changes
are on the open string coupling and the T-dual radius (denoted as $\bar
G_o$
and  $\bar
R_3$, respectively) as
\begin{equation}
\bar R_3 = \frac{\alpha'_{\rm eff}}{R_3}, \ \ \bar G^2_o = \frac{G^2_o 
\sqrt{\alpha'_{\rm eff}}}{R_3} = \frac{R_2}{R_3},
\end{equation}
where we have used Eq. (\ref{eq:pr}) for the last equality in the second
equation. We will not address potential subtleties(for examples, 
see \cite{klem,kawt,dangkone,dangktwo}) related to the
T-dualities of decoupled theories here. 

\subsection{Compactification on a magnetic circle}

	The compactification of the deformed OM theory on a magnetic
circle\footnote{The
compactification along $x^5$-direction seems to give the usual 
(1 + 4)-dimensional NCYM. The boost is now an internal momentum and
appears to be invisible to the NCYM. For this reason, we consider here
only the compactification along either the $x^3$ or $x^4$ direction.} 
(say, a circle along the $x^3$ direction) follows the usual story 
to give a (1 + 4)-dimensional NCYM with a rank-2 noncommutative matrix.
A further T-duality, say, along the $x^2$ direction gives (1 +
3)-dimensional NCYM with again a rank-2 noncommutative matrix. The
V-duality for either (1 + 4)-dimensional or (1 + 3)-dimensional NCYM as 
discussed in~\cite{CLW}, as we will see, follows from that for OM theory.

	To begin with, consider M-theory again on $R^{10}\times S^1$
(but now the circle is along the $x^3$ direction) with M5 branes
wrapping on this circle. The scaling limits for the bulk moduli for OM
theory is again given by Eq. (\ref{eq:comsl}). The relation between
M-theory and IIA now implies that the above OM theory is equivalent to a
decoupled theory of IIA theory as
\begin{eqnarray}
&&\alpha' = \epsilon' \alpha'_{\rm eff}, \ \ g_s = \epsilon'^{1/2} 
\left(\frac{R_3}{l_{\rm eff}}\right)^{3/2}, \ \  \cos\alpha =
\epsilon', \ \ \gamma = \tilde v  
\epsilon',\nonumber\\
&&g_{\mu\nu} = \eta_{\mu\nu}~(\mu,
\nu = 0, 1, 2), \ \ g_{ij} = \epsilon'^2 \delta_{ij}~ (i, j =  4, 5),\ \
g_{m n}  = \epsilon'^2 \delta_{mn}~ (m, n = {\rm transverse}),\nonumber\\
&& 2\pi\alpha' B = - \epsilon' \tan\alpha \sinh\gamma d\tilde t \wedge    
d\tilde x^4, - \epsilon'^2 \tan\alpha \cosh\gamma
d\tilde x^4 \wedge d\tilde x^5,
\label{eq:ncymdl}
\end{eqnarray}
where we have set the new scaling parameter $\epsilon' = \epsilon^{1/2}$
and 
\begin{equation}
\alpha'_{\rm eff} = \frac{l^3_{\rm eff}}{R_3}.
\label{eq:a}
\end{equation}

	The above scaling limit is the one just for (1 + 4)-dimensional
NCYM as expected. We actually have a family of (1 + 4)-dimensional NCYM 
theories which are related to the one corresponding to $\tilde v = 0$ by
a Galilean boost given in (\ref{eq:ngb}). We can see this easily by
computing the Seiberg-Witten open string moduli, for the metric,
\begin{equation}
G_{\alpha\beta} = \left(\begin{array}{ccccc}                           
                                - (1 - \tilde v^2)&0&0&0&- \tilde v\\
                                  0&1&0&0&0\\
                                  0&0&1&0&0\\
                                  0&0&0&1&0\\
                                 -\tilde v&0&0&0&1\end{array}\right),
\end{equation}
and for the nonvanishing noncommutative parameter
\begin{equation}
 \Theta^{45} = 2\pi\alpha'_{\rm eff}.
\end{equation}
In the above, $\alpha, \beta = 0, 1, 2, 4, 5$.

	The above open string metric is related to the one with $\tilde
v = 0$ by the Galilean boost (\ref{eq:ngb}) and the noncommutative
parameter remains intact under this boost as it should be. We therefore
shows that the V-duality for NCYM as discussed in~\cite{CLW} can also be
obtained from that for OM theory. A T-duality of the above along, say,
the $\tilde x^2$-direction will give a family of (1 + 3)-dimensional
NCYM theories which are again related to the one with $\tilde v =0$ by
the V-duality discussed in~\cite{CLW}. But we have it here from that for 
OM theory. This T-duality does not touch the noncommutative parameter
but reduces trivially the open string metric to a rank-4 one through
dropping $G_{2\beta}$ and $G_{\alpha 2}$. The changes are on the closed
string coupling, the gauge
coupling and the radius of circle in the T-dual, i.e., the
$x^2$-direction. If we denote the T-dual closed string coupling,
coordinate radius
and the gauge coupling as $\bar g_s, \bar R_2$ and $\bar g^2_{\rm YM}$, 
respectively (the original ones as $g_s, R_2, g^2_{\rm YM}$), we have
first
\begin{equation}
\bar R_2 = \frac{\alpha'_{\rm eff}}{R_2}, \ \  \bar g_s = g_s
\epsilon'^{1/2} \frac{\sqrt{\alpha'_{\rm eff}}}{R_2} = \epsilon'
\frac{R_3}{R_2},
\label{eq:tpr}
\end{equation}
where we have used $g_s$ and $\alpha'_{\rm eff}$ given in Eqs. 
(\ref{eq:ncymdl}) and (\ref{eq:a}) in the last equality of the second
equation above.  The Yang-Mills coupling for the (1 + 4)-dimensional
NCYM can be calculated as
\begin{equation}
g^2_{\rm YM} = (2\pi)^2 g_s \sqrt{\alpha'} \left(\frac{\det G}
{\det(g + 2\pi\alpha' B)}\right)^{1/2} = 4\pi^2 R_3.
\end{equation}
The Yang-Mills coupling for the (1 + 3)-dimensional NCYM is related to
the above as
\begin{equation}
\bar g^2_{\rm YM} = \frac{\bar g_s}{2\pi g_s \sqrt{\alpha'}} g^2_{\rm YM}
= 2\pi \frac{R_3}{R_2},
\end{equation}
where we have used the relation between $\bar g_s$ and $g_s$ as given above.

\sect{Conclusion}

	We show in this paper that the (M5, M2, M$2'$, MW) bound state of
M theory constructed recently in \cite{Berg3} through uplifting 
the (D4, D2, D$2'$, D0) bound state of IIA theory to eleven dimensions
 is actually related
to the (M5, M2) bound state of M-theory by a finite Lorentz boost along a 
M5 brane direction perpendicular to the M2-brane. This motivates us to
consider the V-duality for OM theory. We find indeed that the V-duality 
holds for OM theory. The meaning of this is that the allowed 
non-trivial \footnote{By 'non-trivial', we mean that the boost does 
not belong to the symmetry group of the original decoupled OM
theory. For example, we don't consider the boost along $x^1$ or $x^2$ 
directions.} spacetime boost on a given OM theory is a Galilean one, 
and such related OM theories are also physically equivalent. 
Everything here goes along the same line as what has
been discussed in~\cite{CW,CLW} for NCOS and NCYM.

	We further show that upon compactification on either an electric
or a magnetic circle (as well as T-dualities), the V-duality for OM
theory gives that for either NCOS or NCYM. This enforces our conclusion
made in~\cite{CLW} and to be reported in detail in a forthcoming
paper~\cite{clwone} that V-duality holds in general for little m-theory
without gravity.  It is also interesting to seek connections between
V-duality discussed here and the recent work on the non-relativistic
closed string theory (NRCS) and its strong coupling dual, i.e., the Galilean
membrane theory\footnote{By definition, the NCRS and its strong coupling
dual require the directions of the
respective background field to be compactified but not the presence of 
the corresponding background brane.} ~\cite{gomo,dangktwo,dangkone}. We wish to
report this in~\cite{clwone}.

\section*{Acknowledgements}
The work of RGC and NO was supported in part by the Japan Society for the
Promotion of Science and in part by Grants-in-Aid for Scientific Research
Nos. 99020 and 12640270, and by a Grant-in-Aid on the Priority Area:
Supersymmetry and Unified Theory of Elementary Particles. JXL acknowledges
the support of U.S. Department of Energy. The research of YSW was supported 
in part by National Science Foundation under Grant No. PHY-9907701.



\begin{thebibliography}{99}
\bibitem{SW}N. Seiberg and E. Witten, JHEP {\bf 0009} (1999) 032,
  hep-th/9908142.
\bibitem{SST}N. Seiberg, L. Susskind and N. Toumbas, JHEP {\bf 0006} (2000) 021, hep-th/0005040.
\bibitem{Gop1} R. Gopakumar, J. Maldacena, S. Minwalla and A. Strominger,
    JHEP {\bf 0006} (2000) 036,  hep-th/0005048.
\bibitem{CHL}C.S. Chu, P.M. Ho and M. Li, Nucl.Phys. {\bf B574} (2000) 275,
    hep-th/9911153.
\bibitem{Chak}S. Chakravarty, K. Dasgupta, O.J. Ganor and G. Rajesh,
        Nucl. Phys. {\bf B587} (2000) 228,  hep-th/0002175.
\bibitem{Berg}E. Bergshoeff, D.S. Berman, J.P. van der Schaar and P. Sundell,
       Nucl. Phys. {\bf B590} (2000) 173,  hep-th/0005026.
\bibitem{Kawa}S. Kawamoto and N. Sasakura, JHEP {\bf 0007} (2000)
     014,  hep-th/0005123.
\bibitem{Berm}D.S. Berman and P. Sundell, JHEP {\bf 0010} (2000) 014,
    hep-th/0007052.
\bibitem{Gop2}R. Gopakumar, S. Minwalla, N. Seiberg and A. Strominger,
         JHEP {\bf 0008} (2000) 008, hep-th/0006062.
\bibitem{Berg2} E. Bergshoeff, D.S. Berman, J.P. van der Schaar and P. Sundell,
     Phys. Lett. {\bf B492} (2000) 193,   hep-th/0006112.
\bibitem{Town} J.M. Izquierdo, N.D. Lambert, G. Papadopoulos and P.K.
  Townsend, Nucl. Phys. {\bf B460} (1996) 560. 
\bibitem{Berg3}E. Bergshoeff, R.G. Cai, N. Ohta and P. Townsend, 
         Phys. Lett. {\bf B495}(2000) 201,  hep-th/0009147.
\bibitem{Myer}J.C. Breckenridge, G. Michaud and R.C. Myers, Phys. Rev. D
     {\bf 55} (1997) 6438, hep-th/9611174.
\bibitem{HO}T. Harmark and N. Obers, JHEP {\bf 0003} (2000) 024,
    hep-th/9911169.
\bibitem{CW}G.H. Chen and Y.S. Wu, hep-th/0006013.
\bibitem{CLW} R.G. Cai, J.X. Lu and Y.S. Wu, hep-th/0012239.
\bibitem{clwone} R. G. Cai, J. X. Lu and Y. S. Wu, to appear.
\bibitem{Perry} M. Perry and J.H. Schwarz, Nucl. Phys. {\bf B489} (1997) 47,
    hep-th/9611065.
\bibitem{Howe}P.S. Howe, E. Sezgin and P.C. West, Phys. Lett. {\bf B399} 
    (1997) 49, hep-th/9702008; Phys. Lett. {\bf B400} (1997) 255, 
   hep-th/9702111.
\bibitem{Harm}T. Harmark, Nucl. Phys. {\bf B593} (2000) 76, hep-th/0007147.

\bibitem{klem} I. Klebanov and J. Maldacena, ``1 + 1 dimensional NCOS
and its $U(N)$ gauge theory dual'', hep-th/0006085.

\bibitem{kawt} T. Kawano and S. Terashima, ``S-duality from OM-theory'',
Phys. Lett. {\bf B495} (2000) 207, hep-th/0006225.

\bibitem{dangkone} U. Danielsson, A. Goijosa and M. Kruczenski, ``
Newtonian Gravitons and D-brane Collective Coordinates in Wound 
String Theory'', hep-th/0012183.

\bibitem{dangktwo}  U. Danielsson, A. Goijosa and M. Kruczenski, ``
IIA/B Wound and Wrapped'', JHEP {\bf 0010}, 020 (2000), hep-th/0009182.

\bibitem{gomo} J. Gomis and H. Ooguri, ``Non-relativistic Closed String
Theory'', hep-th/0009181.

\end{thebibliography}
\end{document}